\documentclass[reqno,12pt]{article}
\usepackage{amsmath, amsgen,amstext,amsbsy,amsopn,amsthm, amssymb}
\newcommand{\version}{November 9, 2004}
\usepackage{graphicx}

\usepackage{latexsym}

\theoremstyle{definition}

\newcommand{\be}{\begin{equation}}
\newcommand{\ee}{\end{equation}}
\newcommand{\beq}{\begin{equation*}}
\newcommand{\eeq}{\end{equation*}}
\newcommand{\bea}{\begin{eqnarray}}
\newcommand{\eea}{\end{eqnarray}}
\newcommand{\beqa}{\begin{eqnarray*}}
\newcommand{\eeqa}{\end{eqnarray*}}

\newcommand{\R}{\mathbb{R}}

\newcommand{\C}{\mathbb{C}}

\newcommand{\h}{\mathcal H}

\newcommand{\bh}{\mathcal B({\mathcal H})}
\newcommand{\m}{\mathcal M}
\newcommand\mo{\m(\mathcal O)}

\newcommand{\bho}{\mathcal B({\mathcal H}_{1})}
\newcommand{\bht}{\mathcal B({\mathcal H}_{2})}

\def\mfr#1/#2{\hbox{$\frac{{#1} }{ {#2}}$}}

\begin{document} 
    \markboth{\scriptsize{JY \version}}{\scriptsize{JY \version}}
    \title{The Role of Type III Factors\\ in Quantum Field 
Theory\footnotemark
    \footnotetext{Lecture given at the von Neumann Centennial 
Conference, Budapest, October 15--20, 2003}}

\author{Jakob Yngvason\\ \normalsize\it Institut f\"ur
Theoretische Physik, Universit{\"a}t Wien,\\ \normalsize\it 
Boltzmanngasse 5, 1090
Vienna, Austria\\\normalsize\it  and \\\normalsize\it  
\normalsize\it 
 Erwin Schr{\"o}dinger 
Institute for Mathematical Physics,\\ \normalsize\it Boltzmanngasse 9,
1090 Vienna, Austria\\
{\small\rm Email: yngvason@thor.thp.univie.ac.at}} 
\date{\version}
\maketitle
\begin{abstract}
One of von Neumann's motivations for developing the theory of operator
algebras and his and Murray's 1936 classification of factors was the
question of possible decompositions of quantum systems into
independent parts.  For quantum systems with a finite number of
degrees of freedom the simplest possibility, i.e., factors of type I
in the terminology of Murray and von Neumann, are perfectly adequate.
In relativistic quantum field theory (RQFT), on the other hand,
factors of type III occur naturally.  The same holds true in quantum
statistical mechanics of infinite systems.  In this brief review some
physical consequences of the type III property of the von Neumann
algebras corresponding to localized observables in RQFT and their
difference from the type I case will be discussed.  The cumulative
effort of many people over more than 30 years has established a
remarkable uniqueness result: The local algebras in RQFT are
generically isomorphic to the unique, hyperfinite type ${\rm III}_{1}$
factor in Connes' classification of 1973.  Specific theories are
characterized by the net structure of the collection of these
isomorphic algebras for different space-time regions, i.e., the way
they are embedded into each other.  \end{abstract}

John von Neumann was the father of the Hilbert space formulation of
quantum mechanics \cite{vN} that has been the basis of almost all
mathematically rigorous investigations of the theory to this day. We
start by recalling the main concepts and explaining some notations.

The {\em observables} of a quantum system are self adjoint operators
$A=A^*$ on a complex, separable Hilbert space $\h$.  For mathematical
convenience we consider only observables in the the algebra $\bh$ of
bounded operators on $\h$.  If $\psi$ is a normalized vector in $\h$,
i.e., $\langle\psi,\psi\rangle=1$, where $\langle\cdot,\cdot\rangle$
denotes the scalar product on $\h$, then the {\it expectation value}
of an observable $A$ is given by
\be\omega_{\psi}(A)=\langle\psi,A\psi\rangle.\ee This is a positive,
linear functional of $A$ with $\omega_{\psi}({\bf 1})=1$ and
$\omega_{\psi}$ is referred to as the {\em state} defined by $\psi$.
More general states are given by {\em density matrices} $\rho$:
\be\omega_{\rho}(A)={\rm
tr}(\rho A)=\sum_{i}\lambda_{i}
\langle\psi_{i},A\psi_{i}\rangle\ee
where $\lambda_{i}\geq 0$ with $\sum_{i}\lambda_{i}=1$ are the 
eigenvalues and $\psi_{i}$ the normalized eigenvectors of the 
positive trace 
class operator $\rho$.

If ${\mathcal S}\subset \bh$ then its {\em commutant} is defined as
\be{\mathcal S}'=\{B\in \bh:\, [A,B]=0 \text{ for all } A\in{\mathcal 
S}\}.\ee This is always a subalgebra of $\bh$  and if $\mathcal S$ is 
$*$-invariant then so is $\mathcal S'$. Moreover, it is closed in the topology defined
by the states. A {\it von Neumann algebra} $\m$ is a $*$-subalgebra of
$\bh$ that is equal to its double commutant, i.e.,
\be\m=\m^{\prime\prime}.\ee
A basic lemma of von Neumann says that this is equivalent to the 
algebra being closed in the topology defined by the states.

In a series of four papers \cite{MN, MN2, MN3, MN4} Murray and von 
Neumann studied
special von Neumann algebras, called {\em factors}.  These are the
algebras $\m$ such that \be\m\vee \m'\equiv \{AB:\, A\in 
\m,\,B\in\m'\}^{\prime\prime}=\bh, \ee i.e., $\bh$ is ``factorized'' 
into $\m$ and its commutant, $\m'$. This is equivalent to
\be \m\cap\m'=\C{\bf 1},\ee i.e., the center of $\m$ contains
only multiples of the identity operator.  The mathematical problem
addressed by Murray and von Neumann was to classify all possibilities
for such algebras.

This problem is motivated by questions of mathematical nature but also
``several aspects of the quantum mechanical formalism strongly suggest
the elucidation of this subject'' \cite{MN}. One of these aspects is
the division of a quantum system into two independent subsystems. In
the simplest case this is achieved as follows. The Hilbert space is
written as a tensor product,
\be\h=\h_{1}\otimes\h_{2}.\ee
The observables of one system {} correspond to the (self adjoint)
elements of $\m=\bho\otimes{\bf 1}$ and the other to the commutant
$\m'={\bf 1}\otimes\bht$.  The observables of the total system are
thus factorized:
\be\bh=\bho\otimes\bht.\ee
Such a factorization of the observable algebra of the total system 
based on a tensor product factorization of the underlying Hilbert 
space is, in the terminology of Murray and von Neumann, the 
 {\bf Type I} case. It is characterized by the existence 
 of {\it minimal projections} in $\m$:
If $\psi\in \h_{1}$ and $E_{\psi}=|\psi\rangle\langle\psi|$ is the 
 corresponding projection on the one-dimensional subspace of $\h_{1}$ 
 generated by $\psi$, then
\be E=E_{\psi}\otimes{\bf 1}\in\m\ee
 is a {\it minimal projection}, i.e., 
 it has no proper subprojections in $\m$.

  The other extreme is the {\bf Type III} case. Here for every
  non-zero {} projection $E\in\m$ there exists a $W\in \m$ which maps
  $E \h$ isometrically onto $\h$, i.e.  \be W^*W={\bf 1}\, ,\quad
  WW^*=E.\ee Von Neumann himself knew only sporadic examples of type
  III factors and he may, in fact, not have been fully aware of their
  significance. He was apparently more attracted by the {\bf Type II}
  case that lies between the two extremes just described: A type ${\rm
  II}$ factor has no minimal projections, but every {} non-zero
  projection $E$ has a subprojection $F<E$ that is finite in the sense
  that $F'<F$, $W^*W=F$, $WW^*=F'$ implies $F'=F$. This case, although
  very interesting from a mathematical point if view, is less
  important in quantum field theory than the other two and will not be
  discussed further here.
  
 A finer classification of type III factors, based on Tomita-Takesaki
 modular theory \cite{Tomita}, was pioneered by A. Connes
 \cite{Connes}.  There is a continuum of nonequivalent types, denoted
 ${\rm III}_{\lambda}$ with $0\leq \lambda\leq 1$, and the case ${\rm
 III}_{1}$ turns out to be of particular importance for relativistic
 quantum physics.\footnote{Examples of factors of type ${\rm
 III}_\lambda$ with $0<\lambda<1$ were given already in 1967 by Powers
 \cite{Powers} and of type ${\rm III}_1$ by Araki and Woods in 1968
 \cite{ArakiWoods}.} The characteristic feature of type ${\rm III}_1$ 
 is that the spectrum of the Tomita-Takesaki modular groups is the 
 whole of $\R$. Among the factors that are generated by an
 increasing family of finite dimensional subalgebras (such factors are
 called {\it hyperfinite}) there is, up to equivalence, only {\it one}
 factor of type ${\rm III}_{1}$.  This uniqueness result is due to U.
 Haagerup \cite{Haagerup}.
 
The only case a student of quantum mechanics is likely to encounter in
standard texts is the type I case and this is, in fact, perfectly
adequate as long as one deals only with systems with a finite number
of degrees of freedom.\footnote{This can be attributed to von
Neumann's theorem on the uniqueness of representations of the
canonical commutation relations for systems with a finitely many
degrees of freedom.} Hence it is a natural to ask whether the other
cases are not just mathematical curiosities that a physicist need not
bother about.

It will be argued below that it {\it is} important to keep the other
possibilities in mind, otherwise one can be led to false physical
conclusions. Investigations on the foundations of {\em relativistic
quantum field theory} (RQFT) revealed already more than 40 years ago
that the algebras generated by observables that are localized in
bounded regions of space-time cannot be type I and that they are, in
fact, generically of type III. This is far from being obvious but
turns out to be a consequence of general physical requirements of
RQFT. Another case where type I is excluded is the statistical physics
of infinite systems with nonzero density \cite{ArakiWoods1}.  In the
present review we shall not discuss the latter but focus on the
situation in RQFT.
 
 The physical difference between the type I situation and the actual
 state of affairs in RQFT can be illustrated by an instructive {\it
 gedankenexperiment} due to E.\ Fermi \cite{Fermi}. The analysis of
 this experiment drew considerable attention about 10 years ago (see
 \cite{Hegerfeldt} and
\cite{BuchhYng} which contain also references to earlier  literature) 
and
papers on this subject have continued to appear  since then (e.g.,
\cite{Hegerfeldt2, further1, further2}). The  standpoint taken in
\cite{BuchhYng} and which will be explained below is that the 
``causality problems'' associated with the gedankenexperiment are due
to the implicit assumption that the experiment can be described within
a type I framework which is not legitimate in RQFT.

The experiment envisaged in \cite{Fermi} is as follows.  One considers
two atoms, $a$ and $b$, that are separated by a distance $R$.  At time
$t=0$ atom $a$ is in its ground state while atom $b$ is in an excited
state.  Due to the decay of atom $b$ and the emitted radiation
absorbed by atom $a$, the latter will at some time $t>0$ be in an
excited state with nonzero probability.  If the effect of the decay of
atom $b$ does not propagate faster than light, the time $t$ time must
be at least $R/c$, i.e., for $t< R/c$ the state of atom $a$ should be
unchanged.

 Let us consider first an analysis of this experiment within a ``type
 I framework'' which at first sight might look reasonable.  In this
 framework the observable algebra for the atoms plus the radiation
 field is the tensor product of three type I factors: For the atom $a$
 the algebra is $\m_a={\mathcal B}({\mathcal H}_{a})$ on a Hilbert
 space $\h_{a}$, for atom $b$ it is $\m_b={\mathcal B}({\mathcal
 H}_{b})$ on a Hilbert space $\h_{b}$ and for the radiation field it
 is $\m_c={\mathcal B}({\mathcal H}_{c})$ on a Hilbert space
 $\h_{c}$. The observables for the whole system are then
 \be\bh=\m_{a}\otimes\m_{b}\otimes\m_{c}\ee on
 $\h=\h_{a}\otimes\h_{b}\otimes\h_{c}$.

 The initial state of system is
 
\be\label{initial}\omega_{0}=\omega_{a}\otimes\omega_{b}
\otimes\omega_{c}.\ee
 
 Here $\omega_{a}$ is the ground state of atom $a$, $\omega_{b}$ an {}
 excited state of atom $b$ and $\omega_{c}$ the vacuum state of the
 radiation field, i.e. a state without photons.  The state at time
 $t>0$ is now \be\omega_{t}(\cdot)=\omega_{0}(e^{itH}\cdot
 e^{-itH})\ee where $H$ is the Hamiltonian of the total system.  In
 Fermi's original model the information about the spatial situation,
 i.e., the separation $R$ of the atom is encoded in the interaction
 part of the Hamiltonian, but this is not important for the present
 discussion.

Within this setting the probability to find atom $a$ excited at time
$t$ is \be \label{exprob1}P(t)=\omega_{t}(E)\ee with $E=({\bf
1}_{a}-|\psi_{a}\rangle\langle \psi_{a}|)\otimes{\bf 1}_{b}\otimes{\bf
1}_{c}$ where $\psi_{a}$ is the ground state of $a$.  If the signal
from the decay of atom $b$ propagates at most with the speed of light
one might thus expect that $P(t)=0$ for $t<R/c$.  This, however, is at
variance with an analyticity property that follows from a stability
assumption about the Hamiltonian $H$.  Namely, if there is a lower
bound for the energy (i.e., if the Hamiltonian has semibounded
spectrum) then the vector valued function $t\mapsto Ee^{itH}\phi$ can,
for all $\phi\in\h$, be continued to an analytic function of $t$ in
the upper half plane, ${\rm Im}\, t>0$.  From this follows
\cite{Hegerfeldt}:

{\em If $\omega_t(E)=0$ for all $t$ in {\em some} time interval, then 
it is {\em identically} zero.} 

Hence, if atom $a$ becomes excited at all, then this happens {\em
immediately} after $t=0$.  Does this mean that quantum mechanics
predicts faster than light propagation?  The answer to this question
is definitely 'no'.  The lesson is rather: The naive type I ideas
behind the mathematical setup for the gedankenexperiment just
described are flawed \cite{BuchhYng}.  Before discussing this further
we have to introduce some formalism.

{} A proper framework for analyzing causal links in space and time is
relativistic quantum field theory. Its general principles were
formulated in the setting of von Neumann algebras in the 60's by
R. Haag, D. Kastler, H. Araki, H.J. Borchers and others, see
\cite{Haag} and
\cite{BuchhHaag}.
The basic object of any concrete model in RQFT is a {\it net} of von
Neumann algebras, ${\mathcal O}\mapsto \m(\mathcal O)$, on a Hilbert
space $\h$, labeled by subsets $\mathcal O$ of (Minkowski) space-time
$\R^4$. It satisfies
\begin{itemize}
    \item Isotony: ${\mathcal O}_{1}\subset {\mathcal O}_{2}$ implies
    $\m({\mathcal O}_{1})\subset\m({\mathcal O}_{2})$.
    \item Local commutativity: If ${\mathcal O_{1}}$ is space-like 
separated from  ${\mathcal 
O_{2}}$ 
    then\newline 
    $\m({\mathcal O}_{1})\subset\m({\mathcal O}_{2})'$.
    \item Additivity: $
    \left(\bigcup_{x\in \R^4}\m({\mathcal 
O_{0}}+x)\right)^{\prime\prime}=\bh$ for all 
    open ${\mathcal O}_{0}$.
    \end{itemize}
Furthermore, there is a unitary representation $U(x,\Lambda)$ of the 
 Poincar\'e group on $\h$ (here $x\in \R^4$ is a translation of 
space-time and $\Lambda$ a Lorentz transformation) such that
 \be U(x,\Lambda)\mo U(x,\Lambda)^{-1}=\m(\Lambda{\mathcal O}+x).\ee
 The joint spectrum of the generators of $U(x,1)$ (i.e., the 
 {\em energy-momentum spectrum\/}) is a subset of the 
 forward light cone,
 and there is a normalized vector $\Omega\in\h$ ({\em vacuum\/}), 
unique up to a phase 
factor,  that satisfies
\be U(x,\Lambda)\Omega=\Omega\ee
for all Poincar\'e transformations $(x,\Lambda)$.

These general properties define a structure that is  at the 
same time surprisingly
rich and tight. Further natural properties, that can be 
verified in special models, restrict it further and lead in 
particular 
to a remarkable result, obtained through the work of several people 
in the course of more than 30 years :

{\it The local algebras $\mo$ (with $\mathcal O$ open and bounded) 
are, under physically plausible assumptions, the same for all RQFTs, 
namely 
they are isomorphic to the unique hyperfinite type ${\rm III}_{1}$
factor.}

Let us note some consequences of the type III property. First, since
every projection $E\in \mo$ can be written as $WW^*$ with an isometry
$W\in\mo$ the mathematical structure is consistent with the idea that
we can change {\it any} state $\omega$ into an eigenstate of $E$ by a
specific local operation without disturbing it in the causal
complement:
 
 Define $\omega_{W}(A)=\omega (W^*AW)$. Then
 \be \label{prep1}\omega_{W}(E)=\omega(W^*WW^*W)=\omega({\bf 1})=1\ee
 but
 \be 
\label{prep2}\omega_{W}(B)=\omega(W^*BW)=\omega(W^*WB)=\omega(B)\ee
 for $B\in\mo'$.
 
The more refined type ${\rm III}_{1}$ property implies even more:
Every state $\varphi$ can be approximated by $\omega_{W}$ for a {\it
fixed} $\omega$ and {} suitable $W$, arbitrarily well in the norm
topology. In other words: {\it Every state can be prepared locally,
with arbitrary precision, from any other state.}  Closely related is
the fact that a local observable algebra $\mo$ {\it has no pure
states}, i.e., for every $\omega$ there are $\omega_{1}$ and
$\omega_{2}$, different from $\omega$, such that
\be \omega(A)=\mfr1/2\omega_{1}(A)+\mfr1/2\omega_{2}(A)\ee
for all $A\in\mo$. 
On the other hand {\it every state on $\mo$ is a vector state}, 
i.e., for every $\omega$ there is a (non-unique!) $\psi_{\omega}\in 
\h$ such that
\be\label{vector}\omega(A)=\langle 
\psi_{\omega},A\psi_{\omega}\rangle\ee
for all $A\in\mo$.

After this interlude let us now return to Fermi's gedankenexperiment
in light of the differences between the type I and type III
situations. The analysis of
\cite{Hegerfeldt}, that seemed to indicate  superluminal propagation, 
was independent of any details of the model and rested entirely on two
assumptions: Stability, i.e., a lower bound for the energy, and the
formula
\eqref{exprob1} for the excitation probability of atom  $a$. While the
stability assumption is not put to question the  interpretation of 
\eqref{exprob1} as an effect of atom $b$ turns out to be at variance 
with basic principles of RQFT. A correct measure of the effect that
takes these principles into account is perfectly causal.

In order to discuss the experiment within the general framework of
RQFT (but independently of any specific model) one replaces the
observable algebra $\m_{a}$ for atom $a$ by a local algebra
$\m({\mathcal O}_{a})$.  Here ${\mathcal O}_{a}\subset \R^4$ is a of
the form ${\mathcal R}_{a}\times \{0\}$ where ${\mathcal
R}_{a}\subset\R^3$ is a small ball where atom $a$ (more precisely, its
center of mass) is localized.\footnote{Strictly speaking one should
replace $\{0\}$ by a small time interval that takes into account the
possibility that the observables might need a smearing in time to be
well defined.  This would lead to the appearance of a few
$\varepsilon$'s in the formulas below but otherwise this point is not
significant and for simplicity of notation we shall ignore it.} We
also include in the algebra $\m({\mathcal O}_{a})$ the observables of
the electromagnetic field in ${\mathcal O}_{a}$ (i.e., a part of what
was previously denoted $\m_{c}$).  In the same way we replace $\m_{b}$
(and part of $\m_{c}$) by a local algebra $\m({\mathcal O}_{b})$ with
${\mathcal O}_{b}={\mathcal R}_{b}\times \{0\}$ where the ball
${\mathcal R}_{b}$ is a distance $R$ apart from ${\mathcal R}_{a}$.

The first thing to realize is that in order to measure a possible
effect in ${\mathcal R}_{a}$ of the decay of atom $b$ in ${\mathcal
R}_{b}$ one must {\em compare} two states of the total system.  In the
first state, denoted $\omega_{0}$, atom $b$ is at $t=0$ present in
${\mathcal R}_{b}$ and in an excited state.  The second state, denoted
$\omega_{0}^{(0)}$, describes the situation where atom $b$ is in its
ground state (or absent).  In both cases the situation outside
${\mathcal R}_{b}$ is at $t=0$ is given by the ground state of atom
$a$ (with center of mass in ${\mathcal R}_{a}$). In other words: The
states $\omega_{0}$ and $\omega_{0}^{(0)}$ can, at $t=0$, not be
distinguished from the ground state of $a$ by measurements outside of
${\mathcal R}_{b}$ but they differ with respect to measurements inside
${\mathcal R}_{b}$.{}\footnote{It may be necessary to leave the state
unspecified in a small neigborhood of the boundary of ${\mathcal
R}_{b}$, cf. the subsequent discussion of the "split property". But
since this neigbourhood can be arbitrarily small this point is not
relevant for the causality issue.}  With time the states evolve,
respectively, into $\omega_{t}$ and $\omega_{t}^{(0)}$.  (For the sake
of comparison with \cite{Hegerfeldt} we use here the Schr\"odinger
picture.)

What does it mean to say that the state $\omega_{t}$ looks like
$\omega_{t}^{(0)}$ in ${\mathcal R}_{a}$, i.e., the situation in
${\mathcal R}_{a}$ is still unexcited after a time span $t>0$?  The
answer is straightforward: The state $\omega _{t}$ should be
indistinguishable from $\omega_{t}^{(0)}$ for all measurements carried
out in ${\mathcal R}_{a}$ which means that \be
\omega_{t}(A)=\omega_{t}^{(0)}(A)\text{ for all }A\in\m({\mathcal
O}_{a}).\ee The proper measure for the effect of the decay of atom $b$
on the state in ${\mathcal R}_{a}$ is thus obtained by considering the
{\it deviation} of $\omega_t$ from $\omega_{t}^{(0)}$ with respect to
all measurements that can be made in $\mathcal R_a$, i.e.,
\begin{equation}\label{exprob2}D(t)=\sup_{A\in\m({\mathcal
O}_{a});\Vert A\Vert\leq
1}\left|\omega_{t}(A)-\omega_{t}^{(0)}(A)\right|.\end{equation}
    
    At $t=0$ we not only have $D(0)=0$ but, by the definition of
    $\omega_{0}$ and $\omega_{0}^{(0)}$, also
    $\omega_{0}(A)-\omega_{0}^{(0)}(A)=0$ for all $A$ localized at a
    distance $<R$ from ${\mathcal R}_{a}$.  In relativistic quantum
    field theories satisfying the axiom of primitive causality
    \cite{Haagschroer}\footnote{In the present context this means the
    following.  Let $\mathcal O=\mathcal R\times \{0\}$ with $\mathcal
    R\subset \R^3$. Then the von Neumann algebra $\m( \mathcal O)$ is
    the same as the algebra $\m(\hat{\mathcal O})$ where
    $\hat{\mathcal O}$ is the causal closure of ${\mathcal O}$. These
    are the space-time points that are space-like with respect to all
    points that are space-like with respect to $\mathcal O$.} this
    implies that $D(t)$ vanishes for $t<R/c$ where $R$ is the spatial
    distance between ${\mathcal R}_{a}$ and ${\mathcal R}_{b}$. (See
    Figure 1.) Hence there is no causality violation.

    \begin{figure}[htf]
    \hskip1cm
\includegraphics[width=12cm, height=9cm]{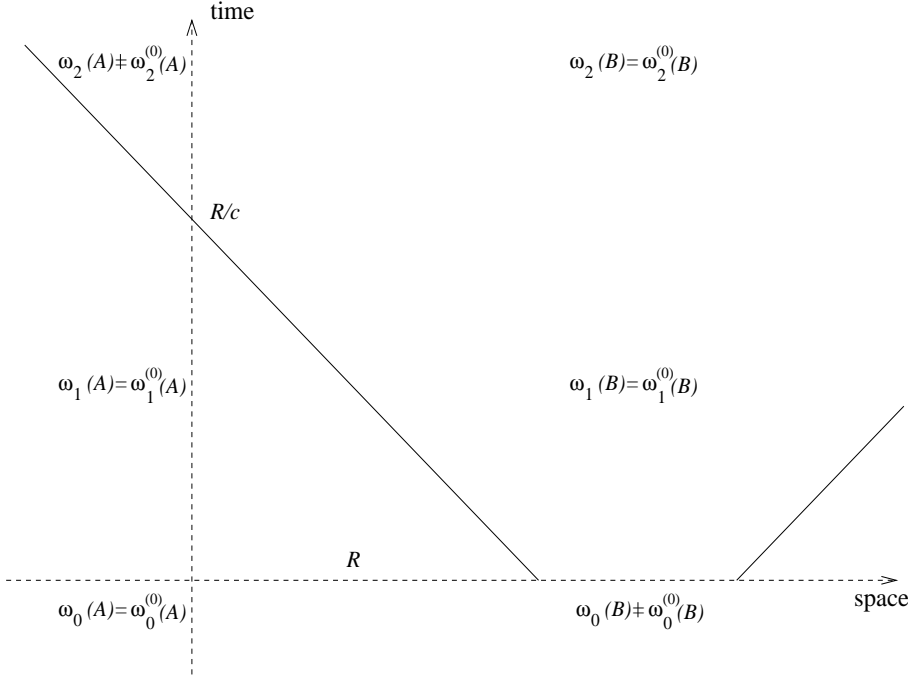}
\caption{Space-time diagram illustrating that 
the situation in $\mathcal R_b$ at $t=0$ does not influence $\mathcal
R_{a}$ for $t<R/c$.}
\end{figure}
       
 In the type I scenario, on the other hand, the excitation probability
 was written as $P(t)=\omega_{t}(E)$ with a projection $E$ and this
 expression cannot vanish on an interval unless it vanishes
 identically.  But since the observable algebra $\m({\mathcal O}_{a})$
 is type III {\em there simply is no projection\/} whose non-vanishing
 expectation value in $\omega_t$ implies a non-vanishing $D(t)$.

  It is instructive to elaborate a little more on this point, in
  particular in view of the fact that $\omega_{0}$ restricted to
  $\m({\mathcal O}_{a})$ is given by a (non-unique) vector
  $\psi\in\h$, cf.  Eq.\eqref{vector}.  The projection on the
  orthogonal complement of $\psi$, $E={\bf
  1}-|\psi\rangle\langle\psi|$, might appear to be a candidate for an
  operator to test the ``excitation'' in region ${\mathcal
  R}_{a}$. This, however, is in error because $E$ is not in
  $\m({\mathcal O}_{a})$ and hence $D(t)$ can well be zero even if
  $\omega_{t}(E)>0$.  The type I situation is different because there
  one can take $E\in \m_{a}$ and $\omega_{t}^{(0)}(E)=0$ for all
  $t$. But if nothing is known about the localization of $E$, even a
  nonvanishing difference $\omega_{t}(E)-\omega_{t}^{(0)}(E)$ does not
  imply any causality violation.
  
  As will be discussed further below, it can in fact be expected that
  $E$ belongs to some $\m({\mathcal O})$ with ${\mathcal O}$ slightly
  larger than ${\mathcal O}_{a}$.  Then causality requires that
  $\omega_{t}(E)-\omega_{t}^{(0)}(E)=0$ for $t<R'/c$ for some $R'$
  slightly smaller than $R$.  This is perfectly compatible with the
  principles of RQFT. It is important to note that the purported
  ``cause'' for $\omega_{t}(E)>0$, i.e., the initial excitation of
  atom $b$, is not present in the state $\omega_{0}^{(0)}$, by
  definition of that state.  But if $\omega_{t}(E)>0$ and
  $\omega_{t}(E)-\omega_{t}^{(0)}(E)=0$ then also
  $\omega_{t}^{(0)}(E)>0$.  Hence a nonvanishing expectation value
  $\omega_{t}(E)$ can certainly not be attributed to a decay of atom
  $b$.
  
The last point becomes even more obvious if we simplify the experiment
by dropping $a$ altogether and take $\omega_{0}^{(0)}$ to be the
vacuum state which is time invariant.  Since $
\omega_{t}(E)-\omega_{0}^{(0)}(E)=0$ for $t<R'/c$ we see that
$\omega_{t}(E)$ is equal to the constant $\omega_{0}^{(0)}(E)>0$ for
$t<R'/c$.  \footnote{Note that, in contrast to the vector valued
function $Ee^{-itH}\phi$, the expectation value $\omega_t(E)$ can {\it
not} be analytically continued to complex $t$ (it involves both
$e^{-itH}$ and $e^{itH}$) and can therefore well be constant on a
whole interval without being constant everywhere.  } The fact that the
vacuum expectation value of any local, nonzero projection is $>0$ is a
consequence of the Reeh-Schlieder theorem \cite{ReehSchlied} which
states that the vacuum vector (or more generally, any analytic vector
for the energy) is cyclic and separating for all $\mo$.  It is clearly
not reasonable to attribute such vacuum fluctuations to some acausal
propagation of signals.

We now leave the discussion of Fermi's gedankenexperiment and 
turn to the general structure of the local algebras of RQFT, starting 
with a brief account of some historic milestones. 

After early discussions on the structure of local algebras 
by Haag and Schroer (1962) \cite{Haagschroer},
Kadison (1963) \cite{Kadison} and Guenin and Misra (1963)
\cite{GueninMisra}, Araki proved in (1963--1964) \cite{Araki1, Araki2,
Araki3} that the local algebras for relativistic free fields are type
III factors.  This establishes the same property also for interacting
fields with local algebras equivalent to those of the free field
(i.e., theories satisfying {\em local normality}).  The relativistic
structure is crucial here because in non-relativistic quantum field
theory the local algebras are type I.

An important general result that {\em almost\/} establishes the type 
III 
property
using only the basic premises of RQFT was obtained by Borchers (1967) 
\cite{Borchers}:  For $\mathcal O\subset \R^4$ open and bounded and 
$\varepsilon>0$ define
\be \label{oeps}{\mathcal 
O}_{\varepsilon}=\bigcup_{|x|<\varepsilon} \left({\mathcal 
O}+x\right).\ee
Then every projection $E\in\m({\mathcal O})$ can be 
written as $WW^*$ with $W\in\m({\mathcal O}_{\varepsilon})$ and
$W^*W={\bf 1}$. 

In 1967 Powers gave explicit examples {} of a continuum of
nonequivalent type III factors, derived from quantum statistical
mechanics. These were the first new examples of type III factor since
von Neumann's paper \cite{MN3}.  Araki and Woods took in 1968 the
first steps towards the finer classification of type III factors and
gave an example of type ${\rm III}_{1}$.  This development culminated
in Connes's classification in 1973
\cite{Connes} based on  Tomita-Takesaki modular 
theory.

For the algebras corresponding to space-like wedges (these are
unbounded domains, generated by two light rays) Driessler
\cite{Driessler1, Driessler2} used modular theory to prove that they
are of type type ${\rm III}_{1}$.  See also \cite{Longo} and
\cite{Borchers98} for more general results along these lines.  
For the wedge algebras the fundamental
papers of Bisognano and Wichmann (1975--1976) \cite{BisWich1,
BisWich2} are very relevant because they allow to identify 
the modular group defined by the vacuum state  with the Lorentz boosts. 
Building on these results but assuming in addition asymptotic scale
invariance, Fredenhagen proved in 1985 \cite{Fredenhag} that also
algebras corresponding to bounded domains (double cones) are type 
${\rm 
III}_{1}$.

In 1987 Haagerup established the uniqueness of the hyperfinite type 
${\rm III}_{1}$ 
factor. Buchholz, D'Antoni, Fredenhagen showed in 1987 
\cite{BuchhDantFred} that natural 
assumptions on the size of sets of states that are approximately
localized in phase space {\it (nuclearity properties}) lead to
hyperfiniteness of the local algebras.  Finally, Buchholz and Verch
\cite{BuchhVerch} in 1996 carried out a comprehensive analysis of the
concept of scaling limits in RQFT and showed that if a model has a
non-trivial scaling limit at all, then the local algebras are type
${\rm III}_{1}$ .
    
 The results of \cite{BuchhDantFred} on the hyperfiniteness are
 related to a more general analysis of the concept of {\it causal
 independence} (also called {\it statistical independence}) that sheds
 further light on the type III property of local algebras. The {}
 review article
\cite{Summers} contains a very thorough discussion of this issue and
we shall here only touch upon a few points.
 Let us first define the 
 relevant concepts.

{\bf Definitions.} 1) A pair of commuting von Neumann algebras,
$\m_{1}$ and $\m_{2}$, in a common $\bh$ is {\it causally
(statistically) independent} if for every pair of states, $\omega_{1}$
on $\m_{1}$ and $\omega_{2}$ on $\m_{2}$, there is a state $\omega$ on
$\m_{1}\vee\m_{2}$ such that
\be\omega(AB)=\omega_{1}(A)\omega_{2}(B)\ee
for $A\in \m_{1}$, $B\in \m_{2}$.  In other words: States can be {\it
independently} prescribed on $\m_{1}$ and $\m_{2}$ and extended to a
common, {\it uncorrelated} state on the joint algebra. (This is really
von Neumann's original concept of independent systems.)

2) We say that the pair $\m_{1}$ and $\m_{2}$ has the {\em split
property} if there is a type I factor $\m$ such that
\begin{equation}\label{split}\m_{1}\subset \m\subset 
\m_{2}'.\end{equation}

Another way to state \eqref{split} is: There is a tensor product
decomposition $\h=\h_{1}\otimes\h_{2}$ such that
\be\m_{1}\subset\bho\otimes{\bf 1}\, ,\quad \m_{2}\subset {\bf
1}\otimes\bht.\ee In the field theoretic context causal independence
and split property for local algebras belonging to space-like
separated domains are essentially equivalent.  More precisely, the
split property clearly implies statistical independence (one can take
$\omega=\omega_{1}\otimes\omega_{2}$), and conversely, statistical
independence implies split property for algebras belonging possibly to
slightly smaller domains than the original algebras \cite{Buchh74}.

Clearly it is desirable to have causal independence (and hence split
property) if $\m_{1}=\m({\mathcal O})$ with $\mathcal O$ some bounded
domain and $\m_{2}$ is generated by the local algebras $\m(\tilde
{\mathcal O})$ with $\tilde {\mathcal O}$ causally separated from
${\mathcal O}$ by some positive distance.  For instance, in Fermi's
gedankenexperiment it is implicitly taken for granted that the state
in $\mathcal R_{b}$ at $t=0$ can be specified independently of that in
the complement of $\mathcal R_{b}$ (with the exclusion of a small
neigborhood of the boundary, cf.  footnote 4). Hence type I factors
play a very important role in RQFT. It is important to note, however,
that the idealization of {\it strict} localization still requires the
local algebras to be type III, even if they can be nested into
commuting type ${\rm I}$ factors, cf.
\eqref{split}, with a slightly ``fuzzy'' localization.  (Note that
$\bho\otimes{\bf 1}$ and ${\bf 1}\otimes\bht$ are not themselves local
algebras although they can be included in local algebras belonging to
slightly larger domains.)

We shall now give an example of a physical application of the split
property.  It was mentioned previously in connection with
\eqref{prep1} --\eqref{prep2} that the type III property has
consequences for the local preparation of states with prescribed
properties.  Combined with the split property a strong version of
\eqref{prep1}--\eqref{prep2} emerges, and since it is easy to prove we
include the proof here.  It is a simplified version of results from
\cite{BuchhDopplLong}, \cite{Wern} and \cite{Summers}.\\

\noindent{\bf Proposition (Strong local preparability).}  {\it For
every state $\omega$ and every bounded $\mathcal O$ there is an
isometry $W\in\m({\mathcal O}_{\varepsilon})$ (with ${\mathcal
O}_{\varepsilon}$ defined by \eqref{oeps}) such that for an arbitrary
state $\varphi$ $$\varphi(W^*AW)=\omega (A)$$ for all $A\in\mo$, but
$$\varphi(W^*BW)=\varphi (B)$$ for all $B\in\m({\mathcal
O}_{\varepsilon})'.$ In words: Every state can be prepared locally
from any other state, with an isometry that depends only on the state
to be prepared.} \\

{\it Proof}: The split property implies that we can write
$\mo\subset\bho\otimes{\bf 1}$, $\m({\mathcal
O}_{\varepsilon})'\subset {\bf 1}\otimes\bht$.  By the type III
property of $\mo$ we have $\omega(A)=\langle\xi,A\xi\rangle$ for
$A\in\mo$, with $\xi=\xi_{1}\otimes\xi_{2}$.  Then
$E=E_{\xi_{1}}\otimes {\bf 1}\in\
\m({\mathcal O}_{\varepsilon})$. By the type III property of 
$\m({\mathcal O}_{\varepsilon})$ 
there is a 
$W\in\m({\mathcal O}_{\varepsilon})$ with $W^*W={\bf 1}$, $WW^*=E$. 
QED\\

In the context of $C^*$ algebras a general result on statistical
independence was obtained by Roos in 1974 \cite{Roos}.  Shortly
afterward the split property was proved to hold for free fields by
Buchholz in 1974 \cite{Buchh74}.  The concept was developed
systematically in \cite{Longoetal} and in \cite{BuchhWich} it was
brought in connection with nuclearity properties that express the idea
that the {\it local} density of states (measured in a suitable sense)
does not increase too fast with the energy.  This line of thought was
developed further in \cite{BuchhDantLong1,BuchhDantLong2} and other
papers, e.g. \cite{BuchhYng2} and \cite{BorchSchu}.

The upshot is that the split property is both physically motivated and
has been proved in models.  It is the basis for the hyperfiniteness of
the local algebras and hence, in connection with \cite{BuchhVerch}, of
the universal structure of the local algebras.\\
 
{\bf Conclusions.} The general postulates of quantum mechanics and
(special) relativity, together with assumptions about the existence of
a scaling limit and bounds on the local density of states, imply a
{\it unique} structure of the local algebras: They are isomorphic to
the unique, hyperfinite type ${\rm III}_{1}$ factor.  Moreover, every
one of these factors is contained in a type I factor that in turn is
contained a local algebra of a slightly larger domain.  This general
structure has important physical implications.  The specific
properties of individual models of RQFT (particle structure,
scattering cross sections, bound states, superselection charges\ldots)
on the other hand, are encoded in the {\it net structure} \be{\mathcal
O}\mapsto \mo\ee i.e.\ in the way these isomorphic type III factors
for different ${\mathcal O}$ are embedded into each other.
 
A challenge for the future, transcending von Neumann's original
program, is to classify all such nets. The recent work of Kawahigashi
and Longo \cite{KaLo}, where this task is carried out for local
conformal nets in two space-time dimensions and with central charge
$c<1$, is an indication that this may not be
an entirely utopian vision.\\

{\bf Acknowledgments} I am indebted to H.J. Borchers and D. Buchholz
for very informative discussions and to the Werner Heisenberg
Institute Munich for hospitality during the preparation of this
lecture.  

\bibliographystyle{amsplain}

\end{document}